# Upper critical field $H_{c2}$ in Bechgaard salts $(TMTSF)_2PF_6$

Ana D Folgueras*[1,2] and Kazumi Maki[3]

Address: [1]Departamento de Física, Universidad de Oviedo, 33007 Oviedo, Spain, [2]Instituto de Ciencia de Materiales de Madrid, C.S.I.C, Cantoblanco, 28049 Madrid, Spain and [3]Department of Physics & Astronomy, University of Southern California, Los Angeles, CA 90089-0484, USA

Email: Ana D Folgueras* - folgueras@gmail.com; Kazumi Maki - kmaki@usc.edu

* Corresponding author

## Abstract

The symmetry of the superconductivity in Bechgaard salts is still unknown, though the triplet pairing has been established by $H_{c2}$ and NMR for $(TMTSF)_2PF_6$. The large upper critical field at T = 0K ($H_{c2}$ ~ 5 Tesla) both for $\vec{H} || \vec{a}$ and $\vec{H} || \vec{b}'$ also indicates strongly the triplet pairing.

Here we start with a low energy effective Hamiltonian and study the temperature dependence of the corresponding $H_{c2}(T)$'s.

The present analysis suggests that one chiral f-wave superconductor should be the most likely candidate near the upper critical field.

**PACS Codes:** 74.70.Kn ; 74.20.Rp; 74.25.Op.

## Introduction

The Bechgaard salt $(TMTSF)_2 PF_6$ is the first organic superconductor discovered in 1980 [1]. Until very recently the superconductivity was believed to be conventional s-wave [2]. More recently the symmetry of the superconductivity has become one of the central issues [3]. The upper critical field at T = 0K in $(TMTSF)_2PF_6$ and $(TMTSF)_2ClO_4$ are clearly beyond the Pauli limit [4-7], suggesting triplet pairing. Recent NMR data [8,9] from $(TMTSF)_2PF_6$ supports triplet superconductivity.

Here we shall first derive $H_{c2}(T)$ for a variety of p-wave and f-wave superconductors [8]. Later, we will discuss the relation between the nuclear spin relaxation rate and the nodal lines.

*Theoretical model*

In the following we shall examine the upper critical field of these superconductors following the standard method initiated by Gor'kov [10] and extended by Luk'yanchuk and Mineev [11] for unconventional superconductors. Also we take the quasiparticle energy in the normal state as in the standard model for Bechgaard salts [2]

$$\xi(k) = v(|k_a| - k_F) - 2t_b \cos(\vec{b}\vec{k}) - 2t_c \cos(\vec{c}\vec{k}) \qquad (1)$$

with $v : v_b : v_c \sim 1 : 1/10 : 1/300$ and $v = v_a$, $v_b = \sqrt{2}t_b b$ and $v_c = \sqrt{2}t_c c$; for example, P. M. Grant [12] gives $v_c \sim 1$ meV, $t_b \sim 26.2$ meV and $t_a \sim 365$ meV.

There are earlier analysis of $H_{c2}$ of Bechgaard salts starting from the one dimensional models [13,14]. However, those models predict diverging $H_{c2}(T)$ for $T \to 0K$ or the reentrance behaviour, which have not been observed in the experiments [4,5]. The one dimensional model, like the one proposed by Lebed [13,15] is valid only when $2t_c < 2.14T_c \sim 3$ K, in Bechgaard salts it is believed that the transfer integral in the $c$ direction is $2t_c \sim 10 - 30$ K while the superconducting transition temperature is $T_c \sim 1.2$ K, so the 1D model is unrealistic. Also, the quasilinear $T$ dependence of $H_{c2}(T)$ in both $(TMTSF)_2PF_6$ and $(TMTSF)_2ClO_4$ is very unusual.

We consider a 3D model, though strongly anisotropic. We start with a continuum model, where the cristal anisotropy is incorporated only through the great anisotropies of the Fermi velocities. We have considered chiral superconductors because these symmetries have been shown to lead to higher $H_{c2}$s. In the absence of an applied magnetic field, we could obtain one of those chiral states as a combination of two different order parameters (with two different transition temperatures), but the external magnetic field breaks the time reversal symmetry, allowing the formation of a chiral state in the superconducting phase (see [16]).

Among the symmetries we have considered, the chiral f'-wave superconductor with $\Delta(\vec{k}) \sim \left(\frac{1}{\sqrt{2}} sgn(k_a) + i \sin \chi_2 \right) \cos \chi_2$, looks most promising, where $\chi_1 = \vec{b}\vec{k}$ and $\chi_2 = \vec{c}\vec{k}$ are $\vec{b}$ and $\vec{c}$ the cristal vectors.

Moreover, if the superconductor belongs to one of the nodal superconductors [17,18] and if nodes lay parallel to $\vec{k}_c$ within the two sheets of the Fermi surface, the angle dependent nuclear spin relaxation rate $T_1^{-1}$ in a magnetic field rotated within the b' - c* plane will tell the nodal directions.

Before proceeding, we show $|(\Delta(\vec{k})|$ of two chiral f-wave superconductors in Fig. 1a) and 1b).

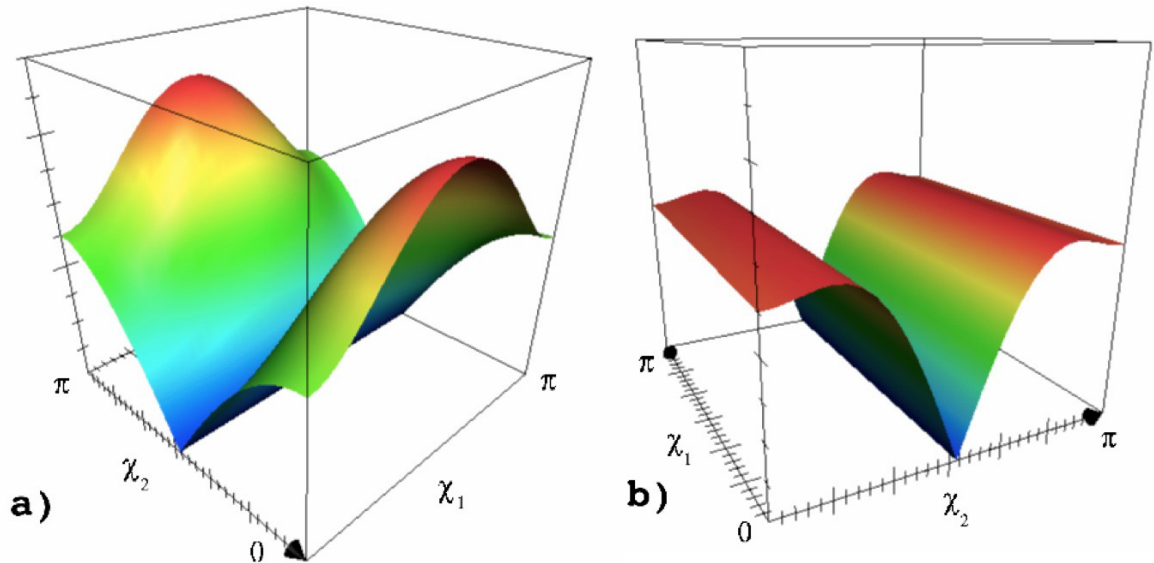

**Figure 1**
**Sketch of the order parameters**. $|\Delta(\vec{k})|$ of chiral f-wave and chiral f'-wave SC are sketched in a) and b) respectively, where $|\Delta(k)| \sim \left[(1+\cos 2\chi_1)(1-\frac{1}{2}\cos 2\chi_2)\right]^{\frac{1}{2}}$ and $|\Delta(k)| \sim \left[(1+\cos 2\chi_2)(1-\frac{1}{2}\cos 2\chi_2)\right]^{\frac{1}{2}}$ for chiral f and chiral f'.

# 1 Results and discussion

## *Upper critical field for* $\vec{H} || \vec{b}'$

In the following we neglect the spin component of $\vec{\Delta}(\vec{k})$. Most likely the equal spin pairing is realised in Bechgaard salts as in $Sr_2RuO_4$ [3]. In this case the spin component is characterized by a unit vector $\hat{d}$. Also $\hat{d}$ is most likely oriented parallel to $\vec{c}^*$. Let's assume $\hat{d} || \vec{c}^*$, though $H_{c2}(T)$ is independent of $\hat{d}$ as long as the spin orbit interaction is negligible. Experimental data from both $UPt_3$ and $Sr_2RuO_4$ indicate that the spin-orbit interactions in these systems are not negligible but extremely small [3]. We consider a variety of triplet superconductors (see some of them in Fig. 1), most of them chiral variants, as we find in general that the chiral variant has larger $H_{c2}$ than the non-chiral one:

*Simple p-wave SC:* $\vec{\Delta}(k) \sim sgn(k_a)$
Following [17,18] the upper critical field is determined by

$$-\ln t = \int_0^\infty \frac{du}{\sinh u}(1 - K_1) \qquad (2)$$

$$-C \ln t = \int_0^\infty \frac{du}{\sinh u}(C - K_2) \tag{3}$$

where

$$K_1 = \langle e^{-\rho u^2 |s|^2}\left(1 + 2C\rho^2 u^4 s^4\right)\rangle \tag{4}$$

$$K_2 = \langle e^{-\rho u^2 |s|^2}\left(\frac{1}{6}\rho^2 u^4 s^{*4} + C\left(1 - 8\rho u^2 |s|^2 + 12\rho^2 u^q |s|^4 - \frac{16}{3}\rho^3 u^6 |s|^6 + \frac{2}{3}\rho^4 u^8 |s|^8\right)\right)\rangle \tag{5}$$

and $t = \frac{T}{T_c}$, $\rho = \frac{v_a v_c e H_{c2}(T)}{2(2\pi T)^2}$, $s = \left(\frac{1}{\sqrt{2}}sgn(k_a) + i\sin\chi_2\right)$, $\chi_2 = \vec{c}\vec{k}$, and $\langle...\rangle$ means average over $\chi_2$. Here $v_a$, $v_c$ are the Fermi velocities parallel to the a axis and the c axis respectively.

Here we assumed that $\Delta(\vec{r})$ is given by [17,18]:

$$\Delta(\vec{r}) \sim \left(1 + C(a^+)^4\right)\rangle \tag{6}$$

where $\rangle = \sum C_n e^{-eBx^2 - nk(x+iz) - \frac{(nk)^2}{4eB}}$ is the Abrikosov state [19], $C_n$ the occupancy of the $n^{th}$ Landau level (we assume there is only one occupied Landau level) and $a^+ = \frac{1}{\sqrt{2eB}}\left(-i\partial_z - \partial_x + 2ieHz\right)$ is the raising operator.

Then in the vicinity of $t \to 1$ we find $\rho = \frac{2}{7\zeta(3)}(-\ln t) = 0.237697(-\ln t)$ and $C = \frac{93\zeta(5)}{647\zeta(3)}\rho$.

For $t \to 0$ on the other hand we obtain

$$\rho_0 = \rho_0(0) = \lim_{t \to 0} \rho t^2 = \frac{v_a v_c e H_{c2}(0)}{2(2\pi T_c)^2} = 0.1583 \tag{7}$$

and $C$ = -0.031. From these we obtain

$$h(0) = \frac{H_{c2}(0)}{\frac{\partial H_{c2}(t)}{\partial t}\big|_{t=1}} = 0.6659 \tag{8}$$

Both $\rho_0(t)$ and $C(t)$ are evaluated numerically and shown in Fig. 2a) and b) respectively. Here $\rho_0(t) = t^2\rho(t) = v_a v_c e H_{c2}(t)/2(2\pi T_c)^2$.

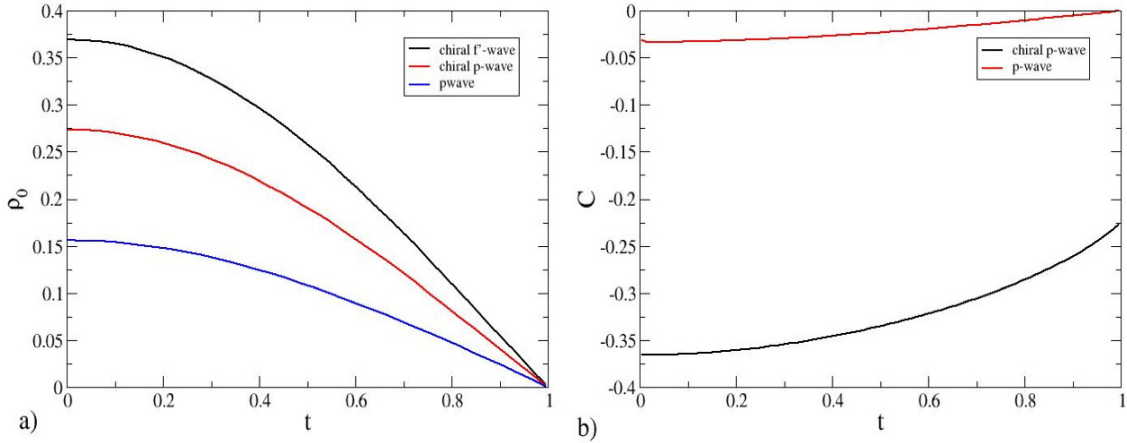

### Figure 2
**Upper critical field for $\vec{H} \parallel \vec{b}'$.** Normalised $H_{c2}(t)$ and $C(t)$ for $\vec{H} \parallel \vec{b}'$ are shown in a) and b) respectively. Here black, red and blue lines are chiral f'-wave, chiral p-wave and simple p-wave respectively. Chiral f-wave has the same $H_{c2}(t)$ as chiral p-wave.

*Chiral p-wave SC:* $\vec{\Delta}(k) = 1/\sqrt{2}\, sgn(k_a) + i\sin(\chi_2)$

Here $\frac{1}{\sqrt{2}} sgn(k_a) + i\sin(\chi_2)$ is the analogue of $e^{i\varphi}$ if in the 3D systems in the quasi 1D system.

For a chiral state the Abrikosov function is written as [20]:

$$\Delta(\vec{r},\vec{k}) \sim \langle (s + Cs^*(a\ )^2) \rangle \qquad (9)$$

where $s = \frac{1}{\sqrt{2}} sgn(k_a) + i\sin(\chi_2)$. Then we obtain eq. 2–3 with

$$K_1 = \langle e^{-\rho u^2 |s|^2}\left(|s|^2 - 2Cs^4\right) \rangle \qquad (10)$$

$$K_2 = \langle e^{-\rho u^2 |s|^2}\left(-s^4 + C|s|^2\left(1 - 4\rho u^2 |s|^2 + 2\rho^2 u^4 |s|^4\right)\right) \rangle \qquad (11)$$

and the same expressions for $t$, $\rho$,...

For $t \to 1$ we find $C = 1 - \sqrt{1.5} = -0.2247$ and $\rho = 0.3838(-\ln t)$.

On the other hand, for $t \to 0$ we obtain $C = -0.3660$ and $\rho_0 = 0.27343$.

From these we obtain $h(0) = 0.71324$. We obtain $\rho(t)$ and $C(t)$ numerically. They are shown in Fig. 2a) and 2b) respectively.

*Chiral f-wave SC:* $\hat{\Delta}(k) \sim \hat{d}s \cos \chi_1$

$H_{c2}(t)$ is determined from eq. 2–3 where now:

$$K_1 = \langle (1 + \cos 2\chi_1) e^{-\rho u^2 |s|^2} \left( |s|^2 - 2\rho u^2 s^4 \right) \rangle \tag{12}$$

$$K_2 = \langle (1 + \cos 2\chi_1) e^{-\rho u^2 |s|^2} \left( -\rho u^2 s^4 + C |s|^2 \left( 1 - 4\rho u^2 |s|^2 + 2\rho^2 u^4 |s|^4 \right) \right) \rangle \tag{13}$$

Here now $\langle ... \rangle$ means the average over both $\chi_1$ and $\chi_2$. As in previous sections, $s = \frac{1}{\sqrt{2}} sgn(k_a) + i \sin(\chi_2)$ ($s$ depends on the direction of the magnetic field). Then it is easy to see that the chiral f-wave SC has the same $H_{c2}(t)$ and $C(t)$ as the chiral p-wave SC, since the variable $\chi_1$ is readily integrated out.

*Chiral f'-wave SC:* $\hat{\Delta}(k) \sim \hat{d}s \cos \chi_2$

Now we have a set of equations similar to the chiral f-wave except $(1 + \cos 2\chi_1)$ in both eqs. 13 has to be replaced by $\frac{4}{3}(1 + \cos 2\chi_1)$. We obtain, for $t \to 1$, $C = -0.2247$ and $\rho = 0.5181(-\ln t)$. On the other hand, for $t \to 0$ we find $C = -0.3660$ and $\rho_0 = 0.3734$.

We show $\rho_0$ and $C(t)$ of the chiral f'-wave in Fig. 2a) and 2b) respectively.

Note that $C(t)$ is the same for three chiral states (chiral p-wave, chiral f-wave and chiral f'-wave) as well as chiral p-wave studied in [20].

Therefore for the magnetic field $\vec{H} \parallel \vec{b}'$, the chiral f'-wave have the largest $H_{c2}(t)$ if we assume $T_c$ and $v$, $v_c$ are the same. Also $H_{c2}(t)$ of these states are closest to the observation.

## Upper critical field for $\vec{H} \parallel \vec{a}$

In this section, we assume the applied magnetic field runs parallel to the direction defined by $\vec{a}$. We calculate the upper critical field in these circumstances for different symmetries of the order parameter, following the same procedure as the one we used in previous section.

*Simple p-wave SC:* $\Delta(\vec{k}) = sgn(k_a)$

The equation for $H_{c2}(t)$ is given by [17,18] and can be written as in eq. 2–3 with:

$$K_1 = \langle e^{-\rho u^2 |s|^2} \left( 1 + 2C\rho^2 u^4 s^4 \right) \rangle \tag{14}$$

$$K_2 = \langle e^{-\rho u^2 |s|^2} \left( \rho^2 u^2 s^4 + C \left( 1 - 8\rho u^2 |s|^2 + 12\rho^2 u^4 |s|^4 - \frac{16}{3}\rho^3 u^6 |s|^6 + \frac{2}{3}\rho^4 u^8 |s|^8 \right) \right) \rangle \quad (15)$$

where $t = \frac{T}{T_c}$, $\rho = \frac{v_b v_c e H_{c2}(t)}{2(2\pi T)^2}$ and $s = (\sin \chi_1 + \iota \sin \chi_2)$ with $\chi_1 = \vec{b}\vec{k}$ and $\chi_2 = \vec{c}\vec{k}$.

Then for $t \to 1$, we find $C = -\frac{93\zeta(5)}{508\zeta(3)}\rho$ and $\rho = \frac{2}{7\zeta(3)}(-\ln t) = 0.2377(-\ln t)$. While for $t \to 0$ $C = \frac{3}{2\beta_0} - \sqrt{\left(\frac{3}{2\beta_0}\right)^2 + \frac{1}{12}} = -0.0170129$ and $\rho_0 = \frac{v_b v_c e H_{c2}(0)}{2(2\pi T_c)^2} = \frac{1}{4\gamma}$, where $\alpha_0 = -\langle \ln |s|^2 \rangle = 0.220051$ and $\beta_0 = -\langle \frac{s^4}{|s|^4}\rangle = \frac{4}{\pi} - 1 = 0.0170$. From these we obtain $h(0) = 0.73673$.

Both $h(t)$ and $C(t)$ are evaluated numerically and we show them in Fig. 3a) and 3b) respectively.

*Chiral p-wave SC:* $\Delta(k) \sim \left( \frac{1}{\sqrt{2}} sgn(k_a) + i \sin \chi_2 \right)$

Now $H_{c2}(t)$ is determined by a similar set of equations as Ec. 10–11. Now, $s = (\sin \chi_1 + \iota \sin \chi_2)$. In particular we find for $t \to 1$ $C = -0.027735$ and $\rho = 0.212598(\ln t)$ while for $t \to 0$ $C = -0.067684$ and $\rho_0 = 0.139672$. We obtain $h(0) = 0.6566$. We show $h(t)$ and $C(t)$ in Fig. 3a) and 3b) respectively.

*Chiral f-wave SC:* $\hat{\Delta}(k) \sim \hat{d}s \cos \chi_1$

Again we use a similar set of equations as Ec. 12–13, with $s = (\sin \chi_1 + \iota \sin \chi_2)$, we find for $t \to 1$ $C = -0.0356236$ and $\rho = 0.2744495(\ln t)$ while for $t \to 0$ $C = 0.066$ and $\rho_0 = 0.1920$ and $h(0) = 0.6997$. Both $h(t)$ and $C(t)$ are evaluated numerically and shown in Fig. 3a) and 3b).

*Chiral f'-wave SC:* $\hat{\Delta}(k) \sim \hat{d}s \cos \chi_2$

Now we find for $t \to 1$ $C = -0.05$ and $\rho = -0.2910(\ln t)$, while for $t \to 0$ $C = -0.1019$ and $\rho_0 = 0.2090$.

We have shown again $h(t)$ and $C(t)$ in Fig. 3a) and 3b) respectively.

Comparing these results with $H_{c2}(T)$ from $(TMTSF)_2PF_6$ and $(TMTSF)_2ClO_4$ [4,5], we can conclude that for both $\vec{H} || \vec{b}'$ and $\vec{H} || \vec{a}$, the chiral f'-wave SC is most consistent with experimental data. In particular these states have relatively large $h(0)$ (see Table 1). On the other hand almost the same $H_{c2}(0)$ for $\vec{H} || \vec{b}'$ and $\vec{H} || \vec{a}$ has to be still accounted.

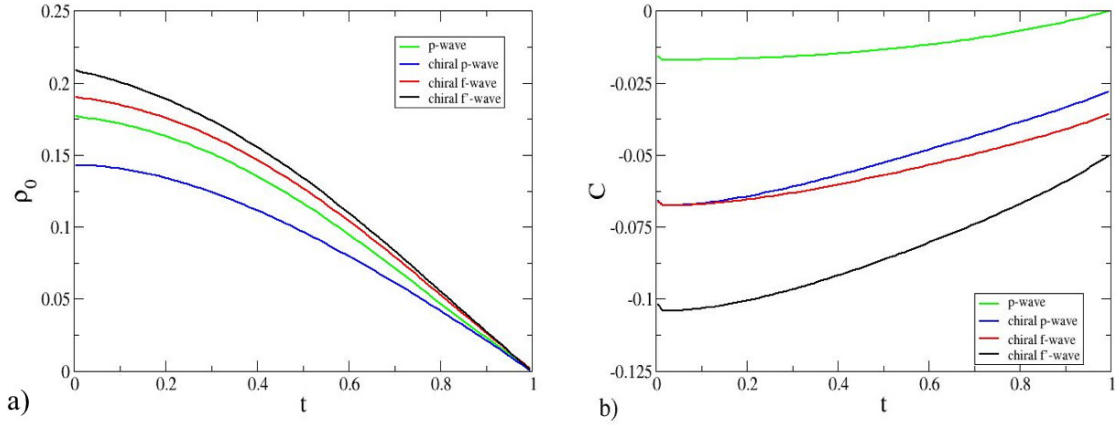

**Figure 3**
**Upper critical field for $\vec{H} || \vec{a}$**. Normalised $H_{c2}(t)$ and $C(t)$ for $\vec{H} || \vec{a}$ are shown in a) and b) respectively. Here black, red, blue and green lines are chiral f'-wave, chiral f-wave, chiral p-wave and simple p-wave respectively.

## Nodal lines in $\Delta(\vec{k})$

We have seen that from the temperature dependence of $H_{c2}(T)$, we can deduce the chiral f-wave and chiral f'-wave superconductors are the most favourable candidates. They have nodal lines on the Fermi surface (i.e. the $\chi_1$ - $\chi_2$ plane), the chiral f-wave SC at $\chi_1 = \pm\frac{\pi}{2}$, while chiral f'-wave SC at $\chi_2 = \pm\frac{\pi}{2}$.

**Table 1: Summary of results.** Here $\rho_0(0) = \frac{v^2 e H_{c2}(0)}{2(2\pi T_c)^2}$ and $h(0) = \frac{H_{c2}(0)}{\frac{\partial H_{c2}(t)}{\partial t}|_{t=1}}$

|   | symmetry | C(0) | C(1) | $-\frac{\partial \rho}{\partial t}|_{t=1}$ | $\rho_0(0)$ | $h(0)$ |
|---|---|---|---|---|---|---|
| $H || b'$ | p-wave | -0.031 | 0 | 0.2377 | 0.1583 | 0.6659 |
|  | chiral p-wave | -0.2247 | -0.3660 | 0.3838 | 0.2734 | 0.71324 |
|  | chiral f'-wave | -0.2247 | -0.3660 | 0.5181 | 0.3734 | 0.72073 |
| $H || a$ | p-wave | -0.017 | 0 | 0.2377 | 0.1751 | 0,7366 |
|  | chiral p-wave | -0.066 | -0.028 | 0.2126 | 0.1396 | 0,6566 |
|  | chiral f-wave | -0.066 | -0.035 | 0.2744 | 0.1920 | 0,6997 |
|  | chiral f'-wave | -0.1019 | -0.05 | 0.2910 | 0.2090 | 0,7182 |

These nodal lines may be detected if the nuclear spin relaxation rate is measured in a magnetic field rotated within the $b'$ - $c^*$ plane.

Following the standard procedure given in [21], the quasiparticle density of states in the vortex state for $T \ll T_c$ and $E = 0$ is given by

$$N\left(0, \vec{H}\right) = \frac{2}{\pi^2} v^2 \sqrt{eH} \left(1 + \cos\theta^2 \sin\chi_{10}^2\right)^{\frac{1}{2}} \qquad (16)$$

where $\chi_{10}$ is the position of the nodal line (i.e. the angle that defines the line on which $\Delta(k) = 0$). So for the chiral f-wave SC we find $\chi_{10} = \frac{\pi}{2}$ and $N(0, \vec{H})$ exhibits the simple angular dependence. On the other hand when nodal lines are on the $\chi_2$ axis, the $\theta$ dependence will be too small to see. Finally this gives

$$T_1^{-1}(\vec{H})/T_{1N}^{-1} = \left(\frac{2}{\pi^2}\right)^2 \vec{v}^2 (eH)\left(1 + \cos\theta^2\right) \qquad (17)$$

for the chiral f-wave SC.

We show the $\theta$ dependence of $T_1^{-1}$ in Fig. 4 for a few candidates. The chiral f-wave SC has the strongest $\theta$ dependence (solid line) while the chiral h-wave SC (dashed line) and the chiral p-wave SC (dotted line) have a similar $\theta$ dependence.

## Conclusion

We have computed the upper critical field of Bechgaard salts for a variety of nodal superconductors with the standard microscopic theory. The results are shown in Fig. 2 and 3. We find:

a) Assuming all these superconductors have the same $T_c$, the chiral f'-wave SC ($\vec{\Delta}(k) \sim \left(\frac{1}{\sqrt{2}} sgn(K_a) + i \sin\chi_2\right)\cos\chi_2$) appears to be the most favourable with largest $H_{c2}$'s for both $\vec{H} || \vec{b}'$ and $\vec{H} || \vec{a}$.

b) However, non of these states exhibit the quasilinear temperature dependence of $H_{c2}(T)$ as observed in [3].

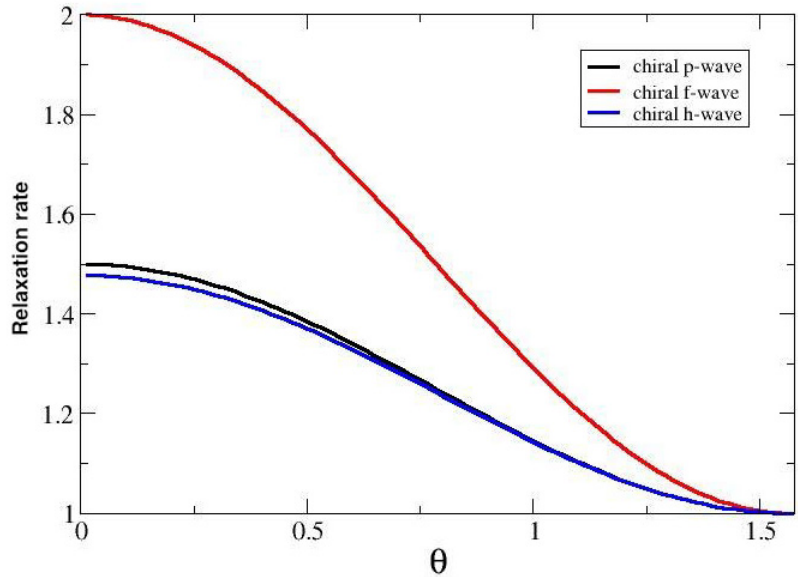

**Figure 4**
**Nuclear spin relaxation rate**. The angle dependent nuclear spin relaxation rate for a few nodal superconductors is shown. Chiral f-wave, chiral h-wave and chiral p-wave are represented in red, blue and black lines respectively. If the nodes lie parallel to $\chi_1$, then it is invisible to NMR, so chiral f'-wave gives the same results as chiral f-wave.

c) Also the present theory predicts $H_{c2}(0) \sim (v_a v_c)^{-1}$ and $(v_b v_c)^{-1}$ for $\vec{H} \parallel \vec{b}'$ and $\vec{H} \parallel \vec{a}$ respectively. This means $H_{c2}(0)$ for $\vec{H} \parallel \vec{a}$ is about 5 time larger than the one for $\vec{H} \parallel \vec{b}'$ contrary to observation.

d) From $H_{c2}(0) \sim 5T$ and $T_c = 1.5$ K we can extract $v^2 = v^2 = \sqrt{v_a v_c} \sim 1.5 10^4 \sim 1.5 10^4$ cm s$^{-1}$, consistent with the known values of $v_a$, $v_c$.

We have also shown that the nodal lines should be visible through the angle dependent $T_1^{-1}$ in NMR with the magnetic field rotating in the c*-b' plane.

### Acknowledgements
We thank S. Brown, P. Chaikin, S. Haas and H. Won for useful discussion. ADF also acknowledges gratefully the discussion with J. Ferrer and F. Guinea. The authors would also like to aknowledge the useful comments of the reviewers during the correction process.

### References
1. Jerome D, Mazard A, Ribault M, Bechgaard K: *J Phys Lett (Paris)* 1980, **47:**L95.
2. Ishiguro T, Yamaji K, Saito G: *Organic Superconductors* Berlin: Springer-Verlag; 1998.
3. Maki K, Haas S, Parker D, Won H: *Chinease J Phys* 2005, **43:**532.
4. Lee IJ, Chaikin PM, Naughton MJ: *Phys Rev B* 2002, **65:**R180502.
5. Oh JI, Naughton MJ: *Phys Rev Lett* 2004, **92:**067001.
6. Clogston AM: *Phys Rev Lett* 1962, **9:**266.

7.  Chandrasekhar BS: *Appl Phys Lett* 1962, **1**:7.
8.  Lee IJ, Brown SE, Clark WG, Strouse MJ, Naughton MJ, Kang W, Chaikin PM: *Phys Rev Lett* 2002, **88**:017004.
9.  Lee IJ, Brown S, Naughton MJ: *J Phys Soc Jpn* 2006, **75**:051011.
10. Gor'kov LP: *Soviet Phys JETP* 1960, **10**:59.
11. Luk'yanchuk I, Mineev VP: *Soviet Phys JETP* 1987, **66**:1168.
12. Grant PM: *J Phys* 1983, **44 C3**:847.
13. Lebed AG: *JETP Lett* 1986, **44**:114.
14. Dupuis N, Montambaux G, Sá de Mello EAR: *Phys Rev Lett* 1993, **70**:2613.
15. Kuroki K, Arita R, Aoki H: *Phys Rev B* 2001, **63**:094509.
16. Sigrist M, Ueda K: *Rev Mod Phys* 1991, **63**:239.
17. Won H, Maki K: *Europhys Lett* 1995, **30**:421.
18. Won H, Maki K: *Phys Rev B* 1996, **53**:5927.
19. Abrikosov AA: *Soviet Phys JETP* 1957, **5**:1174.
20. Wang GF, Maki K: *Europhysic Lett* 1999, **45**:71.
21. Wong H, Haas S, Parker D, Telang S, Vanyolos A, Maki K: **BCS theory of nodal superconductors.** In *Lectures on the Physics of Highly Correlated Electron Systems IX* Edited by: Avella A, Mancini F. AIP; 2005.